\documentclass[eqsecnum,prd]{revtex4}
\usepackage{amsmath} 
\usepackage{hyperref}
\usepackage[latin1]{inputenc}
\usepackage[T1]{fontenc}
\usepackage{mathrsfs}
\usepackage{mathpazo}
\usepackage{graphicx}

\newcommand{\cA}{\mathscr{A}}
\newcommand{\cB}{\mathscr{B}}
\newcommand{\cE}{\mathscr{E}}
\newcommand{\cG}{\mathscr{G}}

\newcommand{\cM}{\mathscr{M}}
\newcommand{\cO}{\mathscr{O}}

\newcommand{\dd}{\mathrm{d}}
\newcommand{\del}{\partial}
\newcommand{\eps}{\epsilon}
\newcommand{\Tr}{\mathop{\mathrm{Tr}}\nolimits}

\newtheorem{theorem}{Theorem}[section]
\newlength{\figurewidth}
\begin{document}

\title{ The static spherically symmetric body in relativistic
elasticity }
 
\author{Jörg Frauendiener}
\email{joergf@cma.uio.no}
\affiliation{Center of Mathematics for
  Applications, University of Oslo, P.O. Box 1053 Blindern, NO-0316
  Oslo, Norway} 
\author{Alexander Kabobel}
\email{alexander.kabobel@uni-tuebingen.de}
\affiliation{Institut für Astronomie und Astrophysik, 
Universität Tübingen, Auf der Morgenstelle 10, D-72076 Tübingen,
Germany}
\date{\today}

\begin{abstract}  
  In this paper is discussed a class of static spherically symmetric
  solutions of the general relativistic elasticity equations. The main
  point of discussion is the comparison of two matter models given in
  terms of their stored energy functionals, i.e., the rule which gives
  the amount of energy stored in the system when it is deformed. Both
  functionals mimic (and for small deformations approximate) the
  classical Kirchhoff-St.~Venant materials but differ in the strain
  variable used. We discuss the behavior of the systems for large
  deformations.
\end{abstract}
\maketitle

\section{Introduction}

Classical elasticity is a phenomenological theory to describe the
properties of solids. It is used heavily in applications such as
structural mechanics and other engineering disciplines. This also
explains the influence of classical elasticity in the formulation of
the popular finite element method for the numerical solution of
partial differential equations (see
e.g.~\cite{hughe2000:the-finite-element-method}). While it is entirely
sufficient to restrict oneself to a classical theory for bulk matter
to describe every-day engineering problems it is nevertheless of conceptual
interest to formulate a theory of elasticity which is compatible with
the space-time structure established by Einstein's theories of
relativity.

The first attempts to merge elasticity with special relativity go back
to the early 20th century and there have been several other
formulations including
Synge~\cite{synge1959:a-theory-of-elasticity-in-general} and
Rayner~\cite{rayne1963:elasticity-in-general-relativity}. The most
influential work, however, has been the paper by Carter and
Quintana~\cite{cartequint1972:foundations-of-general-relativistic} who
formulated the geometric setting for the theory and derived the basic
field equations. The theory has also been considered from a
field theoretical point of view by Kijowski and
Magli~\cite{kijowmagli1992:relat-elastomechanics}. Recently, the
theory has been analyzed from the point of view of the initial value
problem formulation by Beig and
Schmidt~\cite{beigschmidt2003:_relat_elast}. They showed that the
field equations can be put into a first order symmetric hyperbolic
form and they prove among other things that the Cauchy problem for the
system is well-posed under various circumstances. Based on this
formulation it is shown
in~\cite{beigschmi2006:relativistic-elastostatics-i} that there exist
solutions of the elasticity equations in Newtonian theory and in
special relativity describing elastic bodies in rigid
rotation. In~\cite{anderbeig2006:static-self-gravitating-elastic} it
is proved that there exist solutions of the static elastic equations
for sufficiently weak gravitational
interaction. Losert~\cite{loser2006:static-elastic-shells} analyzes
the case of a self-gravitating elastic spherical shell and shows
existence of solutions in the Newtonian case.

In a series of
papers\cite{karlosamue2003:elastic-stars-in-general,karlosamue2004:elastic-stars-in-general_II,karlosamue2004:elastic-stars-in-general_III}
Karlovini, Samuelsson and Zarroug adopt the formulation of
Carter and Quintana to discuss spherically symmetric equilibrium
configurations and their radial perturbations. They also present an
exact static and spherically symmetric solution with constant energy
density.

Our intention in this paper is to discuss two different equations of
state in the static and spherically symmetric context with two
different materials. In~\cite{beigschmidt2003:_relat_elast} the
familiar Kirchhoff-St.~Venant stored energy functional for
hyper-elastic isotropic materials has been extended to the relativistic
case. Recall that this functional is quadratic in the strain variable
and contains the Lam\'e coefficients as two material constants.
Kijowski and Magli
in~\cite{kijowmagli1992:relat-elastomechanics} use the same
functional. However, they adopt a different definition for their
strain variable which has a non-linear relationship to the strain
used by Beig-Schmidt. Hence, this results in two different stored
energy functionals which have the property that, by construction, they
agree with the classical Kirchhoff-St.-Venant functional in the
non-relativistic, small deformation limit.

The plan of the paper is as follows. In sec.~\ref{sec:prelim} we
provide the necessary background on the formulation of the theory of
relativistic elasticity. The exposition follows that
of~\cite{beigschmidt2003:_relat_elast}. We present the two
formulations by Beig-Schmidt and Kijowski-Magli and point out their
differences. It turns out that the only difference is in the
definition of the strain variable which accounts for the above
mentioned different energy functionals.

In sec.~\ref{sec:spherical_symmetry} we specialize to the static and
spherically symmetric case and derive the equations which govern this
situation. We show that this system of equations has a unique smooth
solution once the central compression of the body has been specified.

Sec.~\ref{sec:numer-model-spher} is devoted to a study of various
models. In order to compare the two energy functionals we consider
various scenarios. We discuss a solid aluminum sphere and a
relativistic highly compact material similar to the nucleonic matter
inside a neutron star, described with both theories as well as with
the classical theory of elasticity. We also discuss how the choice of
different natural states affects the solutions.

\section{Preliminaries}
\label{sec:prelim}
\subsection{Relativistic Elasticity} 
\label{subsec:re}

The relativistic theory of elasticity in the form that we will use in
this work has been described
in~\cite{beigschmidt2003:_relat_elast}. The kinematic structure of the
theory can be formulated as follows. As the basic variable one
considers a (smooth) map
\begin{equation}
f: \cM \to \cB
\end{equation}
from space-time~$(\cM,g)$\footnote{In this paper we use geometric
  units and the conventions of~\cite{penrose84:_spinor_spacet_i}.} to
a 3-dimensional manifold $\cB$, the 'material manifold' or 'body
manifold' or simply `the body'. This is a reference manifold which
carries some additional structure which will be described later. The
body manifold can be interpreted as the collection of all point-like
constituents (baryons) of the actual body. Coordinates on $\cB$ are
labels for each individual `particle' of the body. The map $f$ is a
map from a 4-dimensional to a 3-dimensional manifold so its derivative
$\dd f$ must have a kernel. One requires that $F = \dd f$ has maximal
rank at each point so that this kernel has dimension one and is
spanned by a unit vector field $u^a$.  Using small Latin indices for
tensors on $\cM$ and capital Latin indices for tensors on $\cB$, the
derivative of $f$ may be written as~$F^A_a=\del_af^A$. In local
coordinates $(x^a)$ on~$\cM$ and $(X^A)$ on~$\cB$ the map $f$ is given
by expressions of the form
\begin{equation}
  X^A = f^A(x^a).
\end{equation}

The dynamics of the theory is specified by a Lagrangian density $\rho$
which is regarded as a functional of $f$ and its first derivative
$\dd f$. In addition, it will depend on the metric $g$. Thus, the
action may be written as
\begin{equation}
\label{eq:action}
\cA[f,\dd f;g] = \int\rho[f,\dd f;g]\,\sqrt{-\det g}\,d^4x.
\end{equation}
The Euler-Lagrange equations for this action are
\begin{equation}
\cG_A := \frac1{\sqrt{-\det g}}\del_a\left( \sqrt{-\det g}
  \frac{\del\rho}{\del (\del_af^A)}\right) - \frac{\del\rho}{\del f^A}
= 0.
\end{equation}
From the properties of $f$ one can already derive several useful
consequences. Let $\Omega_{ABC}$ be a 3-form on $\cB$. This 3-form can
be interpreted as defining a measure on $\cB$ which gives to each
subset of $\cB$ the number of particles contained in it. The pull-back
of $\Omega_{ABC}$ to $\cM$ along $f$ is a 3-form $\omega_{abc}$ on
$\cM$ which is dual to a vector field~$\omega^a$. It is clear that
this vector field spans the kernel of $f$ so that it must be
proportional to $u^a$. Hence we have the formulas
\begin{equation}
  \begin{aligned}
    \omega_{abc} &= \Omega_{ABC} F^A_a F^B_b F^C_c =
    \eps_{abcd}\omega^d,\\  
    \omega^a &= n u^a = \frac16 \eps^{abcd}\omega_{bcd}.
  \end{aligned}
\label{eq:omega}
\end{equation}
The proportionality factor $n$ is interpreted as the number density of
particles (baryon density) constituting the body in the state it
acquires when embedded into space-time.

The energy-momentum tensor of the theory is defined as usual by the
variation of the action with respect to the metric
\begin{equation}
T_{ab}  = - 2\frac{\del \rho}{\del
  g^{ab}} + \rho g_{ab}.
\end{equation}
A consequence of the diffeomorphism invariance of the Lagrangian is
(see~\cite{beigschmidt2003:_relat_elast}) that
\begin{equation}
\nabla_bT^{b}{}_a = \cG_AF^A_a
\end{equation}
i.e., that the elastic field equations are satisfied if and only if
the energy-momentum tensor is divergence free. This is not necessarily
the case in other field theories, such as e.g., for the Maxwell field.

The inverse metric $g^{ab}$ defines a contravariant, symmetric and
positive definite 2-tensor $H^{AB}$ on the body by push-forward with
the map~$f$ (the minus sign is due to our signature)
\begin{equation}
H^{AB} = -g^{ab} F_a^AF_b^B.
\end{equation}
This characterizes the current state of the body which can vary due to
the space-time curvature. In order to describe the variation the
conventional way is to compare the actual state with a reference state
that is given a priori as a fixed structure on the body~$\cB$. This
can be done by postulating the esxistence of a (positive definite)
reference metric $\gamma_{AB}$ on the body manifold which
characterizes a `natural' state of the body in which -- by definition
-- there is no strain\footnote{Carter and
  Quintana~\cite{cartequint1972:foundations-of-general-relativistic}
  point out that in the context of neutron stars where the solid
  structure of the material exists only due to the high pressure it is
  useless to specify an undeformed state. They propose a high-pressure
  formulation in which one does not specify a strain-less state but
  for a fixed value of the pressure one specifies a state with
  shear. We will not pursue this formulation further but leave it for
  a separate investigation.}.  The difference
$E^{AB}=H^{AB}-\gamma^{AB}$ between $H^{AB}$ and the inverse
$\gamma^{AB}$ provides a measure of the `size' of the strain on the
body. Equivalently, one may use the linear map $\cE = E^A{}_B =
H^A{}_B - \delta^A_B$ obtained by lowering an index on $E^{AB}$ with
$\gamma_{AB}$. 

Writing $\rho = n e$ where $e$ is the energy per particle then the
second Piola-Kirchhoff stress tensor is obtained as the derivative
\begin{equation}
  \label{eq:piolakirchhoffstress}
  \tau_{AB} =  2\frac{\del e}{\del H^{AB}}.
\end{equation}
Thus, specifying $e$ as a function of the strain provides the
stress-strain relation i.e., the equation of state for the material
under consideration. If the stress tensor $\tau_{AB}$ does not vanish
in the natural state in which there is no strain then one talks about
a \emph{pre-stressed} state, otherwise the state is called
\emph{stress-free} or \emph{relaxed}. We will be concerned only with a
relaxed state. Thus, the energy density has a minimum in the natural
state. For most applications it is enough to assume that the energy
density is at most quadratic in the strain and we will do so
here. Invariance under coordinate transformations in the body implies
that it can depend only on the scalar invariants of~$\cE$ and since we
are in three dimensions those invariants which are at most quadratic
in $\cE$ are $\Tr \cE$ and~$\Tr (\cE^2)$. Thus, the energy
density can be written as
\begin{align*}
  \rho&=n m \left(1+\frac18\left\{p \Tr(\cE^2) + q
      (\Tr\cE)^2\right\}\right) \\
  &= n m \left(1+\frac18\left\{p  (H^{AB}H_{AB}-2H^A{}_A+3) + q  (H^A{}_A-3)^2 
     \right\}\right)
\end{align*}
where $m$ is the rest mass of a particle and $p$ and $q$ are
constants. This is the stored energy functional which is assumed
in~\cite{beigschmidt2003:_relat_elast}. It describes the so-called
Kirchhoff-St. Venant materials. When we refer below to the
Beig-Schmidt (BS) formulation we mean the use of this stored energy
functional. 

The fact that there exists a metric on the body implies that there are
now two 3-forms available: the 3-form $\Omega_{ABC}$ which gives the
number of particles in each sub-domain of the body and the volume form
$V_{ABC}$ induced by $\gamma_{AB}$ which gives the volume of the
sub-domain. Since the two forms must be proportional we have
\begin{equation}
\Omega_{ABC} = n_0 V_{ABC},
\end{equation}
thus defining the \emph{particle density} $n_0$ in the natural
state. This can be used to define the mass density $\rho_0=mn_0$ in
the natural state. Using the `natural' particle density $n_0$ we can
obtain the following formula
\begin{equation}
\omega_{abc} = \Omega_{ABC} F^A_a F^B_b F^C_c = n_0 V_{ABC} F^A_a
F^B_b F^C_c = n \eps_{abcd}u^d = n \eps_{abc}.
\end{equation}
In local coordinates where $\Omega_{ABC} \doteq
\sqrt{\det{\gamma}}\; d^3X$ and $\eps_{abc} 
\doteq \sqrt{-\det{h}}\; d^3x$ with $h_{ab}=g_{ab}-u_au_b$, we have
\begin{equation}
\label{eq:n_n0}
n_0 \sqrt{\det{\gamma}} \det{F} = n \sqrt{-\det{h}}. 
\end{equation}

\subsection{The Kijowski-Magli strain}
\label{subsec:km}

The main difference between the
Beig-Schmidt~\cite{beigschmidt2003:_relat_elast} and
Kijowski-Magli~\cite{kijowmagli1992:relat-elastomechanics}
formulations is the choice of the variable which measures the
deformation. Beig-Schmidt use the difference between the actual and
the relaxed metrics on the body while Kijowski-Magli use a logarithmic
variable. They claim that this variable has better behavior when large
deformations are studied.

With our choice of conventions and notation this variable is
\begin{equation}
S_a{}^b = -\frac12 \log(u_au^b-\gamma_a{}^b),\label{eq:S_def}
\end{equation}
where $\gamma_{ab}=F_a^A F_b^B \gamma_{AB}$ is the pull-back of the
reference metric on $\cB$ to the space-time. Note, that $\gamma_{ab}$
is positive definite so that the tensor inside the parentheses has
only positive eigenvalues and the logarithm is well-defined.

Kijowski-Magli write down an action functional in terms of this
variable. As before, the scalar character of the action implies that
it can depend only on the scalar invariants of $S$ and Kijowski-Magli
assume that it is at most quadratic in $S$. They introduce the
invariants
\begin{equation}
  \label{eq:invariants}
\alpha = \Tr S , \qquad \beta = \frac12 \Tr \tilde S^2  
\end{equation}
where $\tilde S$ is the trace-free part of $S$. Then, they write
the action in the form 
\begin{equation}
  \label{eq:KMaction}
\cA = \int n\left( m + \frac12 A \alpha^2 + B \beta \right)\;
\sqrt{-\det g}\, \dd^4x. 
\end{equation}
Here, we have adapted the formula of Kijowski-Magli somewhat because
we use the particle density $n$ instead of the matter density and
consequently we have to interpret $e$ as the energy per
particle. When we refer to the Kijowski-Magli (KM) formulation we mean
the use of this stored energy functional.

When deriving the equations of motion Kijowski and Magli use familiar
techniques from Lagrangian field theory. However, their
energy-momentum tensor is the \textit{canonical} one and not the
\textit{dynamical} (symmetric) one which is obtained by varying the
action with respect to the metric
(see~\cite{szabados2004:_quasi_local_energ_momen_angul_momen_gr} for a
thorough discussion of this difference). Since we are using the latter
tensor we cannot simply take over the expression of
Kijowski-Magli. Instead, we need to derive this energy-momentum tensor
explicitly as given in appendix~\ref{sec:find-energy-moment}. We
obtain
\begin{equation}
  \label{eq:KM_energy_momentum}
  T_{ab} = \rho\, u_a u_b  + n \alpha A h_{ab} + n B \tilde
  S_{ab} .
\end{equation}

\subsection{Comparison of the two formulations}
\label{sec:comp-betw-beig}

In this section we want to compare the two presented formulations of
relativistic elasticity. We establish that they agree on the
linearized level and show how they differ for large deformations.  In
order to compare these two formulations we introduce the following
variable which measures the deformation from a given state
\[
\eps_a{}^b = -\left( h_{ac} + \gamma_{ac}\right) h^{cb} = -h_a{}^b -
\gamma_a{}^b. 
\]
In terms of $\eps_a{}^b$ we can write the KM
deformation tensor in the form
\[
S = -\frac12 \log(1+\eps).
\]
The BS deformation is $E^{AB} = H^{AB} - \gamma^{AB}$. We can relate
these two difference deformation variables by the following
computation
\begin{equation}
  \begin{aligned}
    E_{ab} &= F^A_aF^B_b E_{AB} = - h^{cd} F^A_aF^B_b \gamma_{AC}
    \gamma_{BD}
    F^D_dF^C_c -  \gamma_{AC} \gamma_{BD} \gamma^{CD} \\
    &= -\gamma_{ac} (\gamma_{bd} + h_{bd} ) h^{cd} = \gamma_{ac}
    \eps_b{}^c.
  \end{aligned}
\end{equation}
It follows that $ E_A{}^A = (H^{AB} - \gamma^{AB}) \gamma_{AB} =
\eps_{ab} h^{ab} = \eps_a{}^a $
and also $E_{AB} E^{AB} = \eps_{ab} \eps^{ab}$. Thus, the BS-energy
density takes the form
\[
\rho_{BS} = n m \left(1 + \frac18 \left\{ p \eps_{ab}\eps^{ab} + q
    (\eps_a{}^a)^2 \right\} \right).
\]
The KM variables $\alpha$ and $\beta$ can be expressed in
terms of $\eps$ as 
well. Thus, e.g., $\alpha$ becomes
\[
\alpha = -\frac12 \log \det(1+\eps)
\]
and, similarly, $\beta$ can be expressed as before in terms of $S$ and
hence in terms of $\eps$. In order to connect with the BS
formulation we expand the energy density up to quadratic terms in
$\eps$. For the expansion of $\alpha$ and $\beta$ we find
\[
S_a{}^b \approx -\frac12 \eps_a{}^b, \qquad \alpha \approx -\frac12
\eps_a{}^a, \qquad \beta \approx \frac18 \eps_{ab} \eps^{ab} -
\frac1{24} (\eps_a{}^a)^2,
\]
so that the energy density of Kijowski-Magli up to second order in
$\eps$ is
\[
\rho_{KM}=n (m + \frac12 A \alpha^2 + B \beta ) \approx 
n \left(m + \frac{A}{8} \left(\eps_a{}^a\right)^2 + \frac{B}{8}
  \eps_{ab}\eps^{ab} -
\frac{B}{24}   (\eps_a{}^a)^2\right).
\]
The expressions for the energy density in the two formulations agree
in this approximation if we put 
\[
mp = B, \qquad mq = A-\frac13 B.
\]
The coefficients in front of the quadratic terms can be related to the
classical elastic constants. Introducing the number density $n_0$ in
the natural state one defines the Lam\'e coefficients $\lambda=n_0mq$
and $2\mu=n_0mp$. Then, $n_0mA$ becomes the bulk modulus $K$ while
$\mu$ is the shear modulus of the material.

Under these circumstances the energy-momentum tensor is given up to
first order terms in $\eps_{ab}$ by 
\[
T_{ab} = \frac{n}{n_0}\rho_0\, u_a u_b - \frac12 \frac{n}{n_0}
\left(2\mu \eps_{ab}  - \lambda \eps_c{}^c \gamma_{ab} 
 \right).
\]
We contrast this with the exact energy-momentum tensors 
\begin{align}
  \text{BS:}\qquad& T_{ab} = \frac{n}{n_0}\rho_0\, u_a u_b + \frac12
  \frac{n}{n_0} \left( 2\mu\, \gamma_{ac} 
    \eps^c{}_b + \lambda\, \eps_c{}^c \gamma_{ab}\right) \label{eq:BS_emt} \\
\text{KM:}\qquad & T_{ab} = \frac{n}{n_0}\rho_0\, u_a u_b  +
\frac{n}{n_0} \left(2\mu \,
  h_{ab}\,\alpha +  \lambda\, S_{ab}\right). \label{eq:KM_emt}
\end{align}
In the case of no deformation i.e., at a point where the body is in
the natural state one has
\[
\gamma_{ab}= -h_{ab}.
\]
The BS-energy-momentum tensor reflects this relationship. It almost
agrees with the linearized energy-momentum tensor except that
$\gamma_{ab}$ appears instead of $-h_{ab}$. Both theories are
quadratic in their respective deformation variables and therefore
describe in some sense a Hookean theory in which stress and strain are
proportional. However, the relationship between the two different
strain variables is highly non-linear. While the two energy-momentum
tensors agree for small strain they disagree heavily for large
deformations. Similarly, the stored energy functionals which give the
energy per particle as a function of strain are completely different
in the two cases when viewed in terms of the strain variable
$\eps_a{}^b$. Thus, the two formulations describe materials with
different equations of state. Both materials behave like the usual
Kirchhoff-St.~Venant materials for small strain, but have a completely
different behavior for large deformations. We want to explore some of
the consequences of these differences in the remainder of this
article.

In the KM formulation the strain variable is defined in terms of the
difference tensors \emph{in space}. This results in a term
proportional to $h_{ab}$ in the energy-momentum tensor, i.e. a term
which is isotropic in space.  In contrast,
in~\cite{beigschmidt2003:_relat_elast} is used the difference between
the actual and the reference state \emph{on the body} as the basic
variable. This results in a term proportional to $\gamma_{ab}$, i.e.,
isotropic on the body. This has the consequence that it is much easier
to describe a fluid as a special case of elastic material within the
KM framework than in the BS case.

\section{Spherical Symmetry}
\label{sec:spherical_symmetry}

Now we specialize to spherical symmetry. We take the space-time metric
to be the general spherically symmetric and static metric
\begin{equation}
g = e^{2\eta}\, d t^2 - e^{2\xi}\,dr^2 - r^2\,(d\theta^2 + \sin^2\theta\,d\phi^2)
\end{equation}
and we assume that the body metric $\gamma$ is spherically symmetric
as well\footnote{Note, however, that this is a priori not at all
  necessary. It could be that the body's natural state is not
  symmetric.}, i.e., when expressed in polar coordinates
\begin{equation}
\gamma_{AB} \doteq e^{2\xi_0}\,dR^2 + R^2\,(d\Theta^2 + \sin^2\Theta\,d\Phi^2).
\end{equation}
Since the geometry at the origin should be regular we need
$e^{2\xi_0}=1$ at the origin, i.e., $\xi_0(0)=0$.
The function $\xi$ and $\eta$ depend on $r$ while $\xi_0$
depends only on $R$. The map $f:\cM \to \cB$ is assumed to be equivariant
and thus without loss of generality it can be expressed as
\begin{equation}
f(r,\theta,\phi) = (F(r),\theta,\phi)
\end{equation}
for some function $F(r)$ with $F(0)=0$. Then, the deformation gradient
is given as 
\begin{equation}
F^A_a \doteq F' \dd r\otimes\del_R + d\theta\otimes \del_\Theta +
d\phi\otimes \del_\Phi .
\end{equation}
Clearly, because of staticity we must have $u^a \doteq
e^{-\eta}\del_t$ and from~(\ref{eq:n_n0})
\begin{equation}
\frac{n}{n_0} = F' \frac{e^{\xi_0} F^2\sin\theta}{e^\xi r^2\sin\theta} =
F'e^{\xi_0-\xi} \frac{F^2}{r^2}.
\end{equation}
The pull-back $\gamma_{ab}$ of the reference metric on $\cB$ is
\begin{equation}
  \label{eq:gamma}
  \gamma_{ab} \doteq  (e^{\xi_0}F')^2 dr^2 + F^2\,
  (d\theta^2 + \sin^2 \theta \,d\phi^2)
\end{equation}

With these formulas and the abbreviations $x=F'e^{-(\xi-\xi_0)}$ and
$y=F/r$ we can compute the deformation tensor
\begin{equation}
  \label{eq:eps}
  \eps_a{}^b \doteq \left(x^2-1\right) dr
  \otimes \del_r + \left(y^2-1 \right)\,
  (d\theta\otimes\del_\theta + d\phi\otimes \del_\phi) 
\end{equation}
Using this variable and the formulas~(\ref{eq:BS_emt})
and~(\ref{eq:KM_emt}) we can find the energy-momentum tensors in both
cases. They are given explicitly in the
appendix~\ref{sec:energy-moment-tens}.

\subsection{The equations}
\label{sec:equations}

The Einstein equations in the spherically symmetric and static case
are well known, see e.g.~\cite{wald84:_gener_relat}. They are
\begin{align}
  \label{eq:G00}
   G_0{}^0 &= e^{-2 \xi} \left\{ \frac1{r^2}
    \left(1-e^{2\xi} \right) - \frac2r  \xi' \right\} = -8\pi 
  \,\rho ,\\
  G_1{}^1 &= e^{-2 \xi} \left\{ \frac1{r^2} 
    \left(1-e^{2\xi} \right) + \frac2r \eta' \right\}
  = 8\pi \, P, \label{eq:G11}\\
  G_2{}^2 &= e^{-2\xi} \left\{ \eta'' + (\eta')^2 -
    \eta' \xi' - \frac1r\left(\eta' - \xi'\right) \right\} =
  8\pi\, Q, \label{eq:G22}
\end{align}
where we have put $T_0{}^0 =\rho$, $P=-T_1{}^1$ and $Q=-T_2{}^2$.
These are three equations for the unknown functions $\xi$, $\eta$ and
$f$ (the function $\xi_0$ which specifies the reference metric is
considered as given). A consequence of the Einstein equations is that
the divergence of the energy-momentum tensor vanishes identically 
\begin{equation}
\nabla_a T^a{}_b=0.
\end{equation}
Under the current conditions this equation has only one non-trivial
component 
\begin{equation}
  P' + \eta' (\rho + P) + \frac2r (P - Q) = 0.
\label{eq:embalance}
\end{equation}
In order to obtain a useful system one replaces the
equation~(\ref{eq:G22}) by~(\ref{eq:embalance}). Furthermore, one 
integrates~(\ref{eq:G00}) by introducing the mass function
\begin{equation}
  \label{eq:mass_function}
  M(r) = 4\pi \int_0^r \rho \bar {r}^2\, d\bar{r}
\end{equation}
or, equivalently, the mean density $w=M(r)/r^3$
to obtain
\begin{equation}
  \label{eq:exp_xi}
  e^{-2\xi} = 1 - 2r^2 w(r).
\end{equation}
Inserting this into~(\ref{eq:G11}) one can solve for $\eta'$ and
insert this into~(\ref{eq:embalance}). Then the following
system of equations is obtained
\begin{align}
  r w' &= -3 w + 4\pi \rho,\\
 r\eta' &= r^2 \frac{4\pi P + w}{1 - 2r^2 w},\label{eq:eta_prime}\\
 r P' &= - r^2\frac{4\pi P + w}{1 - 2r^2 w} (\rho + P) - 2
 (P-Q).\label{eq:P_prime} 
\end{align}
This system is somewhat deceptive, because $\rho$,
$P$ and $Q$ are functions of $F$ and its derivatives. Since they
contain $F$ and $F'$ in a non-linear way the third equation gives a
complicated non-linear equation for $F''$. Equivalently, we will
regard these functions as depending on $x$ and $y$ defined
above. Then $P' = P_xx' + P_y y'$. From their definition we get a
relationship between $x$ and $y$
\begin{equation}
  ry' = x e^{\xi-\xi_0} - y  
\end{equation}
which can be used to substitute for $y'$. With this preparation we
now have the following final system of equations
\begin{equation}
\begin{aligned}
 r w' &= -3 w + 4\pi \rho,\\
 r y' &= \frac{x}{\sqrt{1-2r^2w}} e^{-\xi_0} - y  , \\
 r x' &= \frac{P_y}{P_x}\left(y - \frac{x}{\sqrt{1-2r^2w}} e^{-\xi_0} \right) 
       - r^2 \frac{4\pi P + w}{1 - 2r^2 w} \frac{\rho + P}{P_x} 
       - \frac{2}{P_x} (P-Q).
\end{aligned}\label{eq:ode_system}
\end{equation}
Once a solution of this system is found we can obtain $\eta$ by
integrating~(\ref{eq:eta_prime}), $e^{-2\xi}$ is given
by~(\ref{eq:exp_xi}) and $F$ is found from the definition of $x$. The
functions $\rho$, $P$ and $Q$ are specified by the choice of the
elastic model as functions of $x$ and $y$, while $e^{-\xi_0}$ is any
given function of $r$ characterizing the natural state of the body. It
is only restricted by having the value of unity at the origin.

\subsection{Behavior at the origin}
\label{sec:behaviour-at-origin}

The equations are singular at the origin $r=0$ and it is not a priori
clear whether there exist regular solutions. If there are solutions
which are bounded near the origin then they have specific values there
which can be obtained from the system by putting $r=0$. Then the left
hand sides vanish and from the right hand sides we get
\begin{equation}
  \label{eq:initials}
  w(0) = \frac{4\pi}{3}\rho(0),\quad y(0)=x(0)=:a, \quad P(0)=Q(0).
\end{equation}
This shows us that the only free datum is the value $a$. It
characterizes the volume change of the body at its center. Since the
body should be compressed we assume that $a=\lim_{r\to0}F(r)/r>1$. The
initial value for $w$ can be computed from the expression of $\rho$ in
terms of $x$ and $y$. The third condition states that in the center
the radial and the tangential stresses should be equal and this is a
condition on the matter model which cannot be influenced by specifying
initial conditions. The fact that the central compression is enough to
characterize a solution uniquely is physically reasonable and
corresponds to the fact that a static fluid configuration is uniquely
characterized by the central pressure.

In order to show that with these initial conditions there exist
regular unique solutions we apply the theorem by Rendall and Schmidt
\cite{rendaschmi1991:existence-and-properties-of-spherically}. The
verification of the conditions necessary for that theorem are somewhat
lengthy and we refer the reader to
appendix~\ref{sec:exist-regul-solut}. It follows from this analysis
that for a given value $a$ there exists a unique and smooth solution
of the system of equations~(\ref{eq:ode_system}) in a neighborhood of
the origin.

\section{Numerical modeling of spherical elastic bodies}
\label{sec:numer-model-spher}

\subsection{The models}
\label{subsec:models}

In the rest of this paper we solve the system~(\ref{eq:ode_system})
for several specific matter models. We consider two situations, a sphere
consisting entirely of an ordinary material such as aluminum and a
sphere which consists of material which resembles the neutron star
crust. In both cases we choose the two different energy functionals
corresponding to the BS and KM formulation, respectively. For aluminum
we use the values $\rho_0=2720\, \mathrm{kg/m^3}$, $\lambda=63.3\,
\mathrm{GPa}$ and $\mu=25\, \mathrm{GPa}$.

For the neutron star matter we follow the presentation
in~\cite{haens2001:neutron-star-crusts} where the structure of the
neutron star crust is described in detail. In the crust of a neutron
star the density increases from the outer layer with $10^6
\mathrm{g/cm^3}$ to the inner edge where the density is approximately
$10^{14} \mathrm{g/cm^3}$. While the ground state of the matter is a
lattice which has anisotropic elastic properties it is customary to
approximate it by a homogeneous and isotropic elastic
material. This material is under high pressure and hence it is much
easier to shear it than to compress it\footnote{We wish to thank Curt
  Cutler for clarifications in this respect.}. In fact, the material
is often assumed to be incompressible.  The shear modulus of the
matter in the neutron star crust has been calculated in
e.g.~\cite{strohvan-h1991:shear-modulus-neutron} and we use a value of
$\mu = 10^{26} \mathrm{Pa}$. The fact that the crust material is almost
incompressible means that the bulk modulus $K$ is very large compared
to the shear modulus and hence that also the Lam\'e coefficient
$\lambda$ is very large. We take it here three orders of magnitude
larger than the shear modulus, i.e., $\lambda=10^{29} \mathrm{Pa}$.

In the relativistic theories there is no canonical choice for the
relaxed state of the body. While it seems natural to specify a flat
metric on~$\cB$ this is not necessary. The choice of the relaxed
metric has been discussed in the
literature~\cite{kabobel2001:_symmet_in_relat,%
lukac1977:finite-homogeneous-relativistic}. We follow here
a suggestion 
by Carter~\cite{personalcommunication} according to which one can
obtain the relaxed state by the following procedure. One assumes the
body is heated up until it melts and then one lets it cool down until it
solidifies again. Assuming that the fluid phase is an ideal fluid then
the body settles in a state which can be described by a solution of the
perfect fluid equations. Hence, besides a flat metric we also consider
the spatial metric corresponding to an incompressible fluid with a
constant density $\rho_0$, i.e, we put
\begin{equation}
  \label{eq:xi0}
  e^{-2\xi_0} = 1-\frac{8\pi}{3} \rho_0 r^2.
\end{equation}

As a third formulation we consider the classical
non-relativistic theory of elasticity. The equations for the classical
theory can be obtained from the relativistic equations as the
Newtonian limit, see~\cite{beigschmidt2003:_relat_elast}. The
difference to the relativistic equations is that one puts
$\xi=\xi_0=0$, so that $x=F'$. Furthermore, eq.~(\ref{eq:embalance})
is replaced by
\begin{equation}
  \label{eq:nr_balance}
    P' + \eta' \rho + \frac2r (P - Q) = 0,
\end{equation}
where $\eta'$ is the gravitational force, determined from the equation
\begin{equation}
  \label{eq:eta_prime_nr}
  \eta' = \frac{M(r)}{r^2} = r w.
\end{equation}
The stress components $P$ and $Q$ have the same functional form in
terms of $x$ and $y$ as those for the BS-energy-momentum tensor,
while~$\rho$ is the mass density in the actual state, given by $\rho =
\frac{n}{n_0}\rho_0 = xy^2 \rho_0$.

Hence, the non-relativistic system is
\begin{equation}
\begin{aligned}
 r w' &= -3 w + 4\pi \rho,\\
 r y' &= x - y  , \\
 r x' &= \frac{P_y}{P_x}\left(y - x \right) 
       - \frac{r^2}{P_x} w \rho - \frac{2}{P_x} (P-Q).
\end{aligned}\label{eq:ode_system_nr}
\end{equation}

All the numerical solutions have been obtained using the Runge-Kutta
ODE solver suite provided in MATLAB. The calculation is started with
an initial value $a$ for $x(0)=y(0)$ which is used to calculate the
initial value for $w(0)=4\pi/3 \rho(0)$ from the energy-momentum
tensor. The calculation stops when $P$ vanishes, indicating that the
boundary of the body has been found.

\subsection{Numerical examples}

\begin{figure}[htb]
  \centering
  \includegraphics[width=\figurewidth]{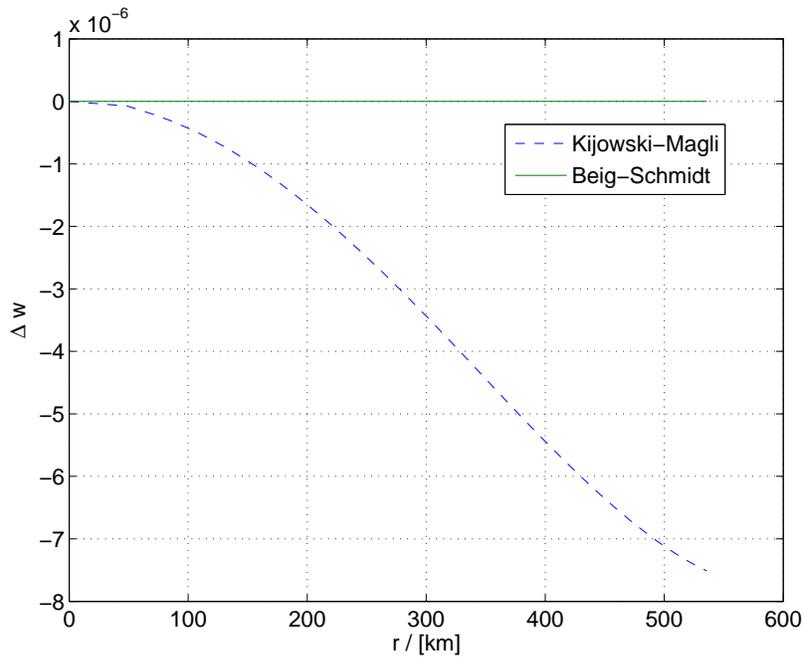}
  \caption{Relative difference between BS resp. KM and the classical
    solution for an aluminum sphere with relative central compression of
    $\delta=0.001$}
  \label{fig:classic}
\end{figure}

We first study the aluminum sphere for the three formulations of
elasticity. Clearly, for small values of the relative central
compression $\delta=a-1$ 
the three formulations should be almost identical. We show in
Fig.~\ref{fig:classic}
the profile of the average density across the sphere for the three
formulations for $\delta=0.001$. In this case we find a sphere with a
radius of $535\,\mathrm{km}$ and a mass of $1.74\times
10^{12}\,\mathrm{kg}$. The figure shows the relative difference
$\Delta w = (w-w_N)/w_N$ between the classical solution $w_N$ and the
BS and KM solutions $w$, respectively. The BS solution is
indistinguishable from the classical solution, the maximum value of
the relative difference being $6\times10^{-13}$, while the KM solution
already indicates its general property: the system is more tightly
bound than in the classical or BS case. Still, in this situation of
small relative central compression the maximal difference is only
$7.5\times 10^{-6}$.

\subsubsection{Aluminum with BS-formulation}
\label{sec:aluminum-with-beig}

\begin{figure}[htb]
  \centering
  \includegraphics[width=\figurewidth]{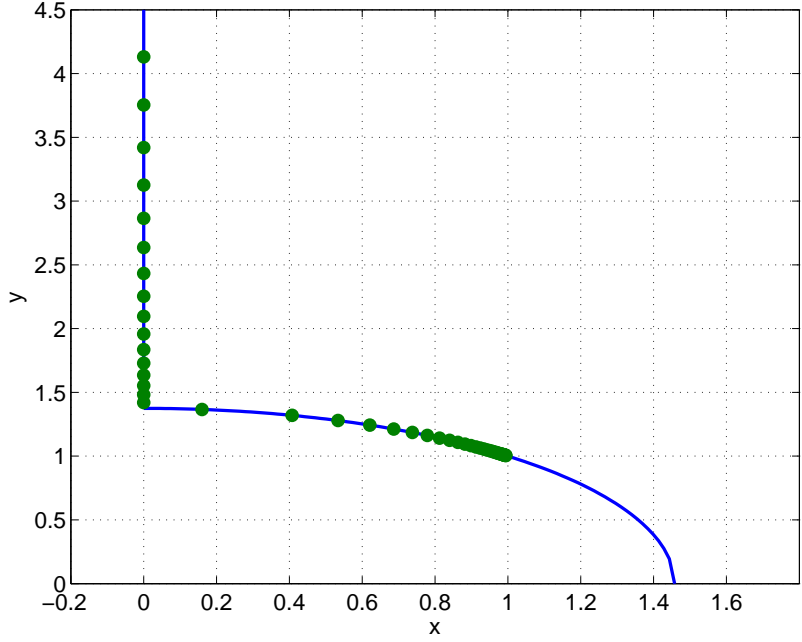}
  \caption{A sequence of pairs $(x,y)$ for relative central compressions
    $0.01\le \delta\le 4$}
  \label{fig:xy_ellipse}
\end{figure}

Let us now look at the BS model in more detail. The radial pressure
is given by~(\ref{eq:BST11})
\begin{equation}
P = \frac12 (xy^2)x^2 \left( (\lambda + 2\mu) (x^2 - 1) 
    + 2 \lambda (y^2 - 1) \right).
\end{equation}
On the boundary of the body, this expression vanishes. This can happen
either when $x=0$ or $y=0$ or if the term in parentheses
vanishes. However, it follows from the equation for $y'$ that as long
as $x$ remains positive we have $y'>0$ for $y=0$ so that $y$ cannot
vanish before $x$ vanishes. Thus, on the boundary we have either $x=0$ or the
pair $(x,y)$ lies on the ellipse defined by 
\begin{equation}
  \label{eq:Pellipse}
   (\lambda + 2\mu) x^2 
    + 2 \lambda y^2 = 3\lambda+2\mu.
\end{equation}
In Fig.~\ref{fig:xy_ellipse}
we show a sequence of such final pairs $(x,y)$ obtained from initial
values $\delta$ in the interval $[0.01,4]$. Obviously, both cases
discussed above can occur. For small relative central compressions the
final pair $(x,y)$ lies on the ellipse and for increasing compression
it moves towards the y-axis until it hits it for an initial value of
$\delta\approx 0.585$. Then it moves along the y-axis for unlimited
values of $y$. The vanishing of $x$ at the boundary means that the
radial distance between two adjacent particles there becomes infinite,
i.e. the body ruptures. Imagine a large elastic sphere without
gravitational self-interaction being compressed so that the central
compression is above the critical value. When gravity is switched on,
the sphere will be divided into a central piece and a shell at the
radius where $x$ vanishes.

The equation for $x'$ in~(\ref{eq:ode_system}) shows that $x$
vanishes with an infinite negative slope because the leading term on
the right hand side goes like $1/x$ near $x=0$. Thus, the solution
becomes singular just at the boundary.

\begin{figure}[htb]
  \centering
  \includegraphics[width=0.4\textwidth]{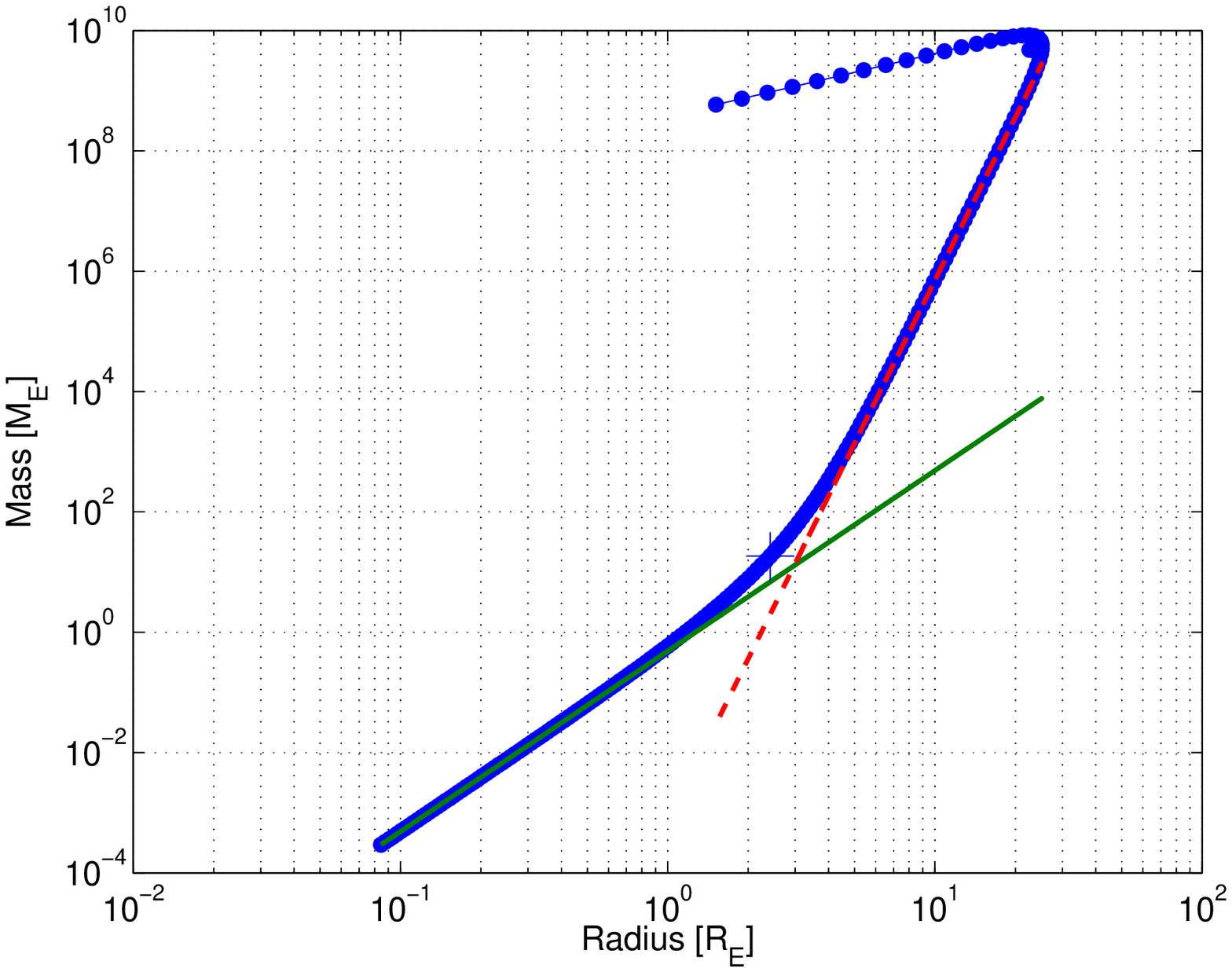}
\qquad
\includegraphics[width=0.4\textwidth]{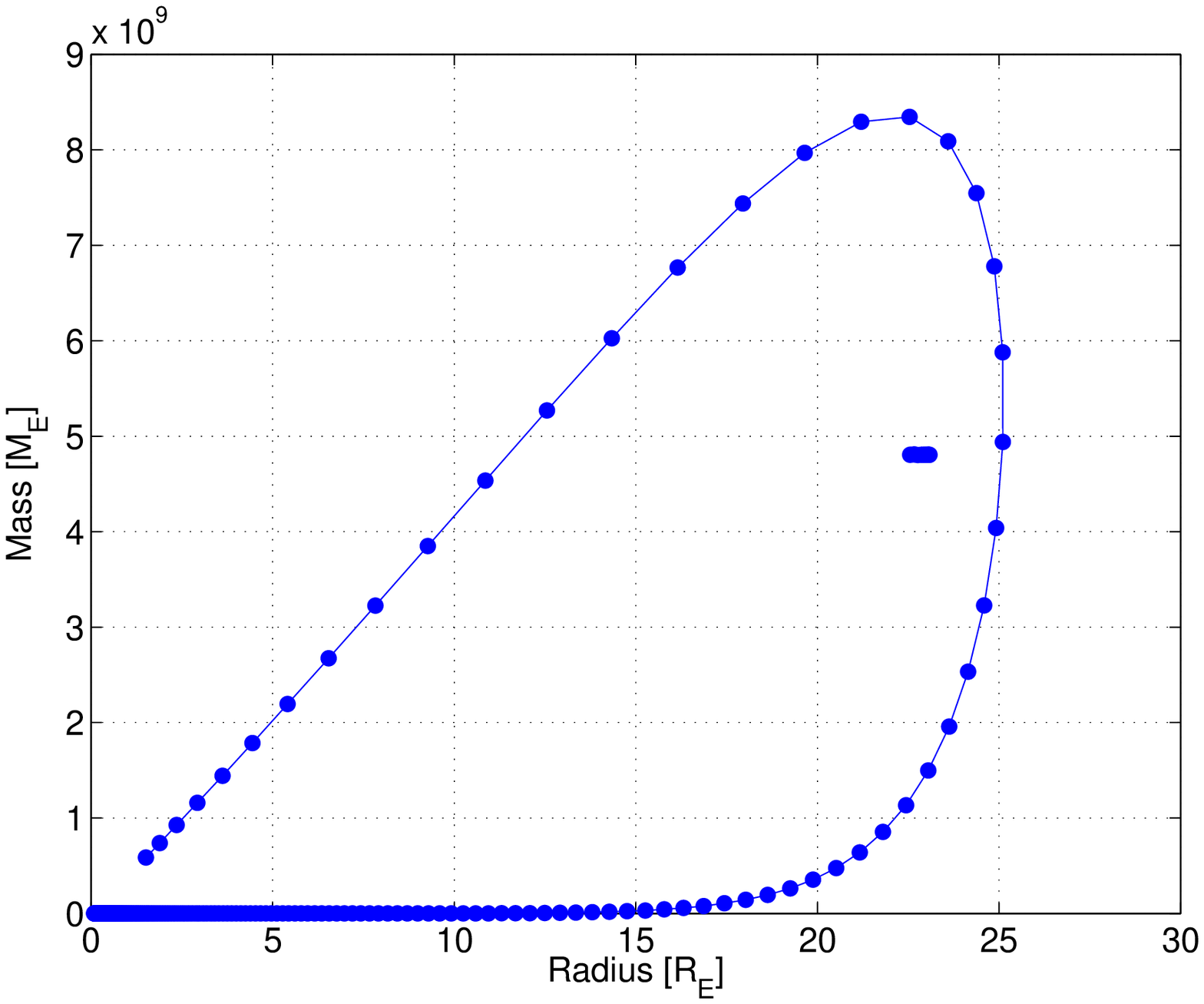}
\caption{Double logarithmic (left) and linear (right) plot of the
  mass-radius diagram for BS aluminum spheres with relative central
  compression $\delta$, $\log_{10}\delta\in [-3,3]$, in units of the
  earth mass and radius, respectively. The solid line is $M\propto
  R^3$ }
  \label{fig:masseradiusBS}
\end{figure}

The two different cases just discussed can also be seen in the
behavior of the mass-radius diagram in Fig.~\ref{fig:masseradiusBS},
where we display radius and mass of the aluminum spheres corresponding
to relative central compressions $\delta \in [10^{-3},10^3]$. We plot
it in double logarithmic and linear axes. The curve shows three
different regimes, the classical one where $M\propto R^3$ (indicated
by the solid line) and an `extreme' regime where $M\propto R^9$,
indicated by the dashed line and finally a `linear' regime with
$M\propto R$ where mass and radius decrease with increasing central
compression. The cross indicates the configuration which is closest to
the critical configuration where the radial strain $x$ vanishes. This
mass-radius diagram should be compared with Figure~1 
from~\cite{karlosamue2004:elastic-stars-in-general_III}. The
similarity of the qualitative behavior is obvious. Karlovini and
Samuelsson argue that the branch from the maximal mass towards zero is
unstable and we do find numerical indications of this here as
well. Increasing the central compression beyond the value needed for
the maximal mass configuration we observe that we can generate the
smaller configurations up to a certain value of $\delta$ depending on
the required precision. Beyond this value the solver suddenly settles
to a solution which yields a configuration in the `eye' inside the
mass-radius diagram. This dot in fact contains nine different
configurations. The location of the `eye' is roughly at the mass
resp. radius for which the radius resp. the mass are maximal on the
curve. The behavior of this system close to the eye should be analyzed
in much more detail using more accurate solution methods.

\subsubsection{Aluminum with KM formulation}
\label{sec:alum-with-kijowski}

\begin{figure}[htb]
  \centering
  \includegraphics[width=\figurewidth]{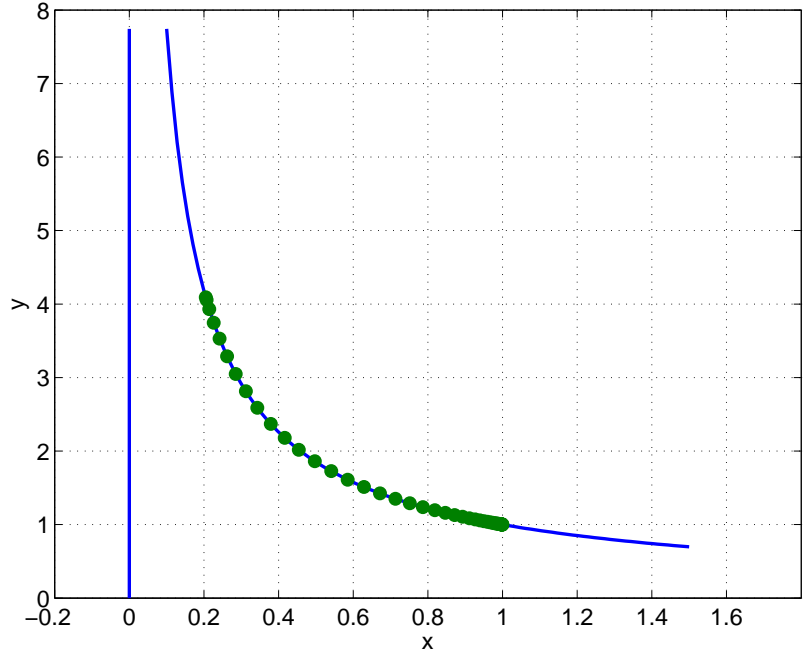}
  \caption{A sequence of pairs $(x,y)$ for relative central compressions
    $0.001\le \delta\le 25$}
  \label{fig:xy_hyperbola}
\end{figure}

\begin{figure}[htb]
  \centering
  \includegraphics[width=\figurewidth]{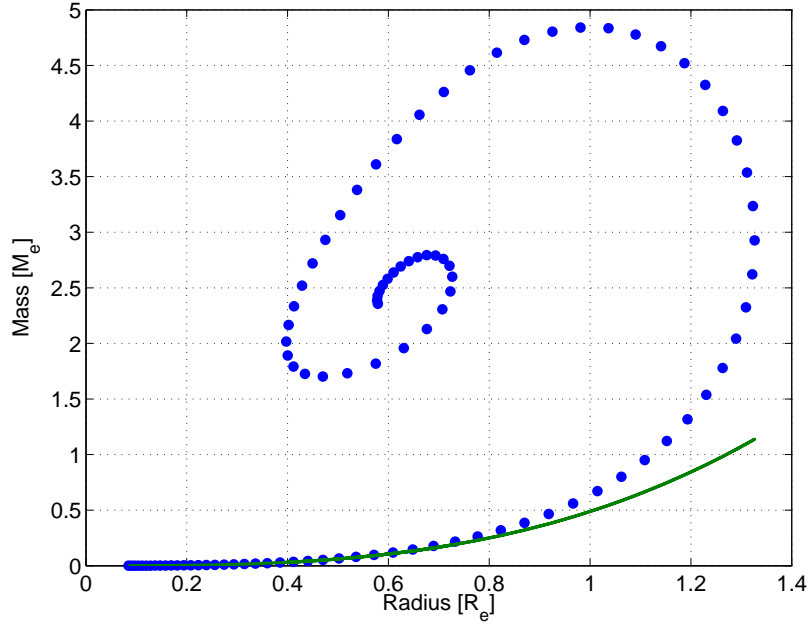}  
  \caption{Mass-radius diagram for KM aluminum spheres with relative central
    compression $\delta$, $\log_{10}\delta\in [-3,3]$, in units of the
    earth mass and radius, respectively. The solid line indicates the
    curve $M\propto R^3$.}
  \label{fig:masseradiusKM}
\end{figure}

In the formulation of Kijowski-Magli the radial pressure $P$ is given
by~(\ref{eq:KMT11})
\begin{equation}
   P = (xy^2) \left( (\lambda + 2\mu)\log x 
    + 2 \lambda\log y \right).
\end{equation}
As before, at the boundary we have either $x=0$ or
\begin{equation}
  (\lambda + 2\mu)\log x + 2 \lambda\log y = 0,
\end{equation}
the case $y=0$ being excluded as before. However, now, a final pair
needs to lie on the curve defined by
\begin{equation}
  \label{eq:Pcurve}
  y = \frac1{x^{\frac{1+\alpha}{2}}}, \qquad \alpha=\frac{2\mu}{\lambda}>0.
\end{equation}
This curve approaches the y-axis but never intersects it. This
indicates that only the case when the final pair lies on the curve
does occur. This is in fact confirmed in Fig.~\ref{fig:xy_hyperbola}
where we show the final pairs $(x,y)$ for aluminum spheres with
relative central compression $\delta \in [0.001,25]$. While in the BS
case the value of $y$ can grow arbitrarily, this is not the case
here. In fact, the numerical investigations show that the exhibited
value of $y\approx4.1$ is the maximal value that $y$ can achieve. This
behavior can be understood when we show the mass-radius diagram for
KM aluminum spheres in Fig.~\ref{fig:masseradiusKM}
which shows a peculiar spiral. The maximal value of $y$ is reached at
the same point as the maximal radius. Thus, it is not possible with the KM
formulation to create arbitrarily large objects. There exists a
maximal mass and a maximal radius for KM aluminum spheres achieved for
different objects and there exists a region where a KM aluminum sphere
of a given radius can have at least four different masses. It looks
like the sequence converges to a limit point. We have not been able to
prove this rigorously.

\subsubsection{The neutron star matter}
\label{sec:neutron-star-matter}

\begin{figure}
  \centering
  \includegraphics[width=\figurewidth]{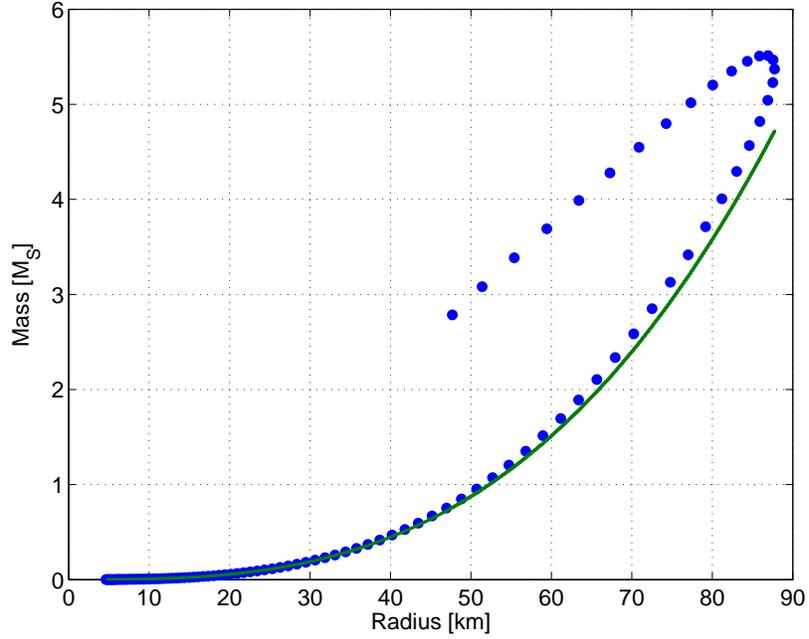}
  \caption{Mass-Radius diagram for BS neutron star like matter with
    relative central compression $\delta$, $\log_{10}\delta \in [-3,0.114]$,
  in units of solar mass and kilometers.}
  \label{fig:NSmasseradiusBS}
\end{figure}

\begin{figure}
  \centering
  \includegraphics[width=\figurewidth]{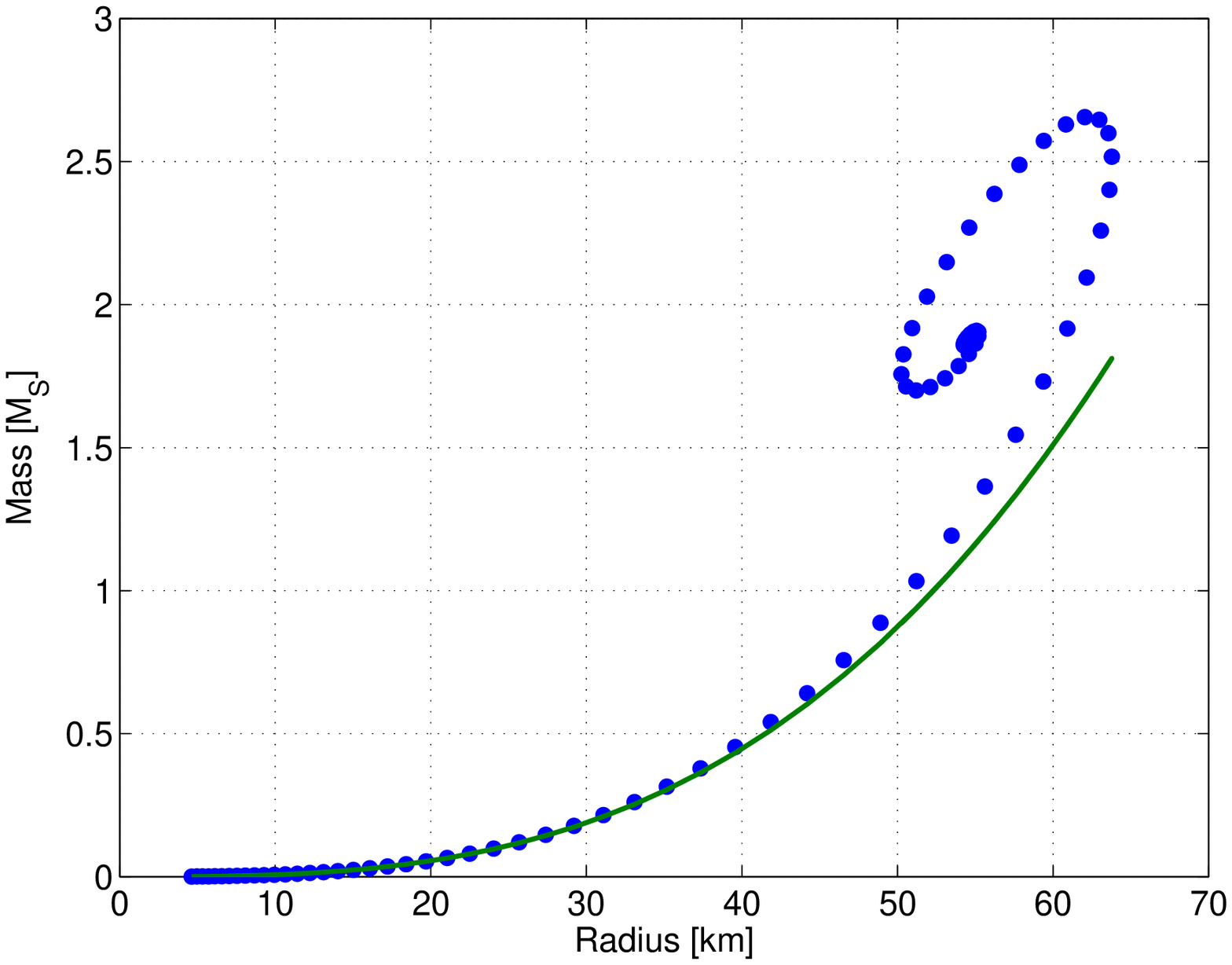}
  \caption{Mass-Radius diagram for KM neutron star like matter with
    relative central compression $\delta$, $\log_{10}\delta \in [-3,3]$,
  in units of solar mass and kilometers.}
  \label{fig:NSmasseradiusKM}
\end{figure}

We have also looked at an exotic material which is somewhat similar to
the nucleonic matter that is assumed to be present in neutron
stars. We show in Fig.~\ref{fig:NSmasseradiusBS}
and in Fig.~\ref{fig:NSmasseradiusKM}
the mass-radius diagrams for the neutron star like matter
distributions with the BS and KM stored energy functionals. In both
cases the diagrams look qualitatively the same as those for aluminum
except that the size of the configurations are orders of magnitudes
different. In the KM case we find a spiral as before while in the BS
case we have the `loop' with a linearly decreasing branch. Again, this
branch seems to be unstable and the final dot in the diagram is the
last for which we could generate a configuration. 

This shows that there is no qualitative difference in the behavior of
aluminum and the exotic matter. This might change if one would use the
high-pressure formulation developed by Carter and 
Quintana~\cite{cartequint1972:foundations-of-general-relativistic}.

\subsubsection{The role of the relaxed metric}
\label{sec:role-relaxed-metric}

\begin{figure}[htb]
  \centering
  \includegraphics[width=\figurewidth]{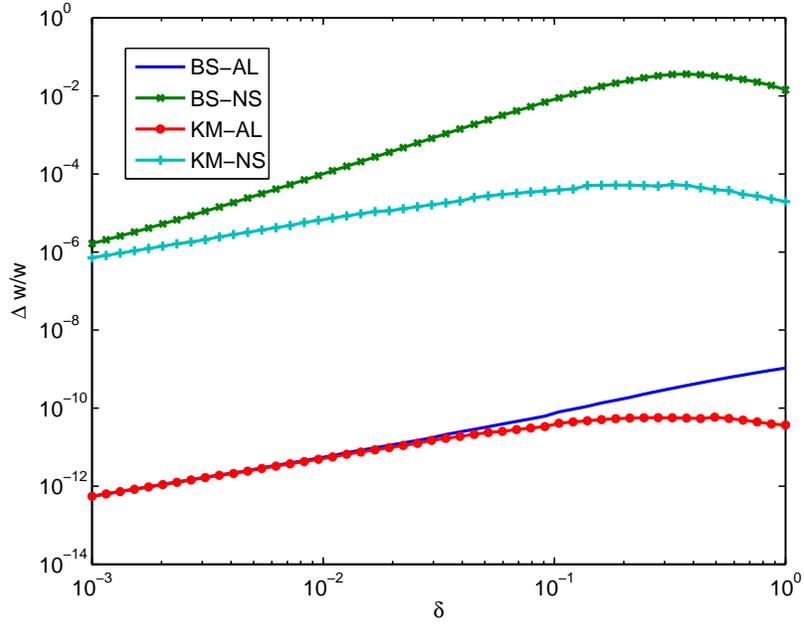}
  \caption{Maximal relative difference in the mean density $w$ between
    calculation with flat and curved metric for the relaxed state for
    given $\delta$ with $\log_{10}\delta \in [-3,0]$. }
  \label{fig:relerr}
\end{figure}

As discussed above we employ two possible choices for the metric of
the relaxed state of an elastic configuration. To compare the two
different scenarios we compute configurations with the same relative
central compression $\delta$ for values of $\delta$ between $10^{-3}$
and $1$ for the two energy functionals and the two possible materials.
In Fig.~\ref{fig:relerr} 
we show the behavior of the maximal absolute value
\[
\frac{\Delta w}{w} = \max_{r} \left|\frac{w_c-w_f}{w_f}\right|
\]
of the relative difference of the mean densities $w_f$ and $w_c$ for
the flat and curved cases resp. as a function of $\delta$.  Obviously,
in the given range of $\delta$ the difference between the two
configurations is almost negligible. The difference is larger for the
exotic neutron star like material than for aluminum. The maximal difference
is reached for the BS-energy functional with roughly 3\%. For
increasing $\delta$ the differences in three cases reach a maximum and
afterwards decrease again. With increasing $\delta$ the elastic energy
in the configuration increases with respect to the gravitational rest
mass energy. Thus, the more the energy of the configuration is
dominated by the elastic energy the smaller is the influence of the
choice of a relaxed state. In any case, what can be learned from
Fig.~\ref{fig:relerr} is, that for practical purposes one can safely
assume that the metric of the relaxed state is flat.

\section{Conclusion}
\label{sec:conclusion}

We have discussed in this work the spherically symmetric body in
relativistic elasticity for two different stored energy
functionals. We find that the BS-functional corresponding to the
classical Kirchhoff-St.~Venant materials and the KM-functional have
entirely different behavior for large deformations even though they
agree for small deformations. The BS-functional gives rise to a
mass-radius diagram which qualitatively is very similar to the one
found by Karlovini and Samuelsson
in~\cite{karlosamue2004:elastic-stars-in-general_III}. They obtain
this diagram for a stiff ultra-rigid equation of state in the
Carter-Quintana high-pressure formulation. They find that the
decreasing branch is unstable. We can confirm this numerically and we
even see indications of another region of configurations. This is an
indication that the BS-functional gives rise to an increasingly stiff
equation of state quite in contrast to the KM-functional for which the
equation of state becomes increasingly soft. The result of this
softness can be seen in qualitatively very different behavior of the
mass-radius diagram which shows a spiral which approaches a limit
point for large deformations.

We looked at these functionals for two different materials, the `every
day' material aluminum and an artificial exotic material. While the
sizes of the individual configurations are very different the
qualitative behavior is very similar in both cases.

In order to analyze in more detail the features in these
configurations and in particular the stability properties of the
different branches of the mass-radius diagrams it might be
advantageous to formulate the problem not as an initial value problem
as we have done here. Instead of specifying the central compression
and integrating outwards to (possibly) find the boundary of a
configuration one would instead set up a boundary value problem on the
body subject to the boundary conditions imposed by the symmetry
requirements in the center and the vanishing of the radial pressure on
the boundary. Steps in this direction have already been made by
Losert~\cite{loser2006:static-elastic-shells}.

Of course, our considerations are to a certain point academic because
any real material will break at already quite moderate deformations
compared to the ones we have used. But we feel that such questions of
principle may shed some light on the differences between the various
possible choices and therefore on the justification of assumptions
made when relativistic elasticity is used for real problems.

\begin{acknowledgments}
  The authors are very grateful to Robert Beig and Bernd Schmidt for
  several very valuable discussions. This work was supported by a
  grant from the Deutsche Forschungsgemeinschaft.
\end{acknowledgments}
\appendix

\section{The symmetric energy-momentum tensor for the Kijowski-Magli
  action} 
\label{sec:find-energy-moment}

The action for the Kijowski-Magli formulation of elasticity
is~(\ref{eq:KMaction}) 
\[
\cA = \int n\left( m + \frac12 A \alpha^2 + B \beta \right)\;
\sqrt{-\det g}\, \dd^4x. 
\]
The energy-momentum tensor for this action is obtained (with our
conventions) by variation of $\cA$ with respect to the inverse metric
\begin{equation}
  T_{ab} = - 2\frac{\del \rho}{\del g^{ab}} + \rho g_{ab}.
\end{equation}
The energy density $\rho$ is specified in terms of the variables
$\alpha = S_a{}^a$ and $\beta = \frac12 S_a{}^b S_b{}^a - \frac16
(S_a{}^a)^2$. The tensor $S_a{}^b$ was defined in~(\ref{eq:S_def})
\begin{equation}
  S_a{}^b = -\frac12 \log \left( u_a u^b - \gamma_a{}^b\right).
\end{equation}
To find the variation in these variables we first need to compute the
variation with $g^{ab}$ in the tensor
\[
K_a{}^b = u_au^b - \gamma_{ac} g^{cb}.
\]
We sketch the calculations here without going too much into the
details. We start with the variation of~(\ref{eq:omega}) to obtain
\[
\delta \omega^d = \frac12\left(g_{ab}\delta g^{ab}\right)\,\omega^d.
\]
Since, $n^2=g_{ab}\omega^a \omega^b$ we can now obtain the variation
in $n$
\[
2\frac{\delta n}{n} = h_{ab} \delta g^{ab}
\]
and with $\omega^a = n u^a$ this yields the variation of $u^b$ and $u_a$ 
\begin{equation}
  \label{eq:1}
  \delta u^b = \frac12 \left(u_cu_d \delta g^{cd}\right) u^b, \qquad
  \delta u_a = \frac12 \left(u_cu_d \delta g^{cd}\right) u_a - g_{ac}
  u_d \delta g^{cd}.
\end{equation}
From this we find
\begin{equation}
  \label{eq:2}
  \delta(u_au^b) = \left(u_cu_d \delta g^{cd}\right) u_au^b - g_{ac}
  u_d u^b \delta g^{cd} = -h_{ac} u_du^b \delta g^{cd}
\end{equation}
and hence
\begin{equation}
  \label{eq:3}
  \delta K_a{}^b = -h_{ac}u_du^b \delta g^{cd} - \gamma_{ac} \delta g^{cb}.
\end{equation}
To compute $\delta \alpha$ we use the formula
\begin{equation}
  \label{eq:4}
  \Tr \log A = \log \det A
\end{equation}
which follows from the corresponding well-known equation
\[
\det \exp B = \exp \Tr B,
\]
valid for any matrix $B$. Thus, we have
\[
\delta \alpha = -\frac12 \delta \Tr \log K = -\frac12 \delta \log\det
K = -\frac12 \frac1{\det K} \delta \det K .
\]
Using the formula
\[
\delta \det K = \det K \Tr(K^{-1}\delta K)
\]
we get the final result
\begin{equation}
  \label{eq:5}
  \delta \alpha = -\frac12 \Tr (K^{-1}\delta K) = -\frac12
  (K^{-1})^b{}_a \delta K^a{}_b.
\end{equation}
The variation of $\beta$ follows from the formula which can easily be
derived  
\[
\delta \Tr f(A) = \Tr (f'(A)\delta A).
\]
We have $\beta = \frac12\Tr (S^2) - \frac16(\Tr S)^2$ so we first compute
\[
\delta \Tr(S^2) = \frac14 \delta \Tr(\log(K)^2) = \frac12 \Tr
(\log(K)K^{-1}\delta K) = - \Tr(S K^{-1} \delta K) =
-S_a{}^b(K^{-1})_b{}^c\delta K_c{}^a.
\]
With this result we obtain 
\begin{equation}
  \label{eq:6}
  2\delta\beta = -\Tr(SK^{-1}\delta K) + \frac13 \Tr S
  \Tr(K^{-1}\delta K) = - \Tr(\tilde SK^{-1}\delta K).
\end{equation}
These formulas can be simplified as follows. Using the fact that
$S_a{}^bu^a=0$ and $K_a{}^b u^a = u^b$ we get
\[
\begin{split}
  (K^{-1})_a{}^b\delta K_b{}^a &= -(K^{-1})_a{}^b( h_{bc}u_du^a \delta
  g^{cd} + \gamma_{bc} \delta g^{ca}) = - (K^{-1})_a{}^b \gamma_{bc} \delta
  g^{ca} = - (K^{-1})_a{}^b (u_b u_c - K_{bc})\delta g^{ca}\\ &= (- u_a
  u_c + g_{ac}) \delta g^{ca} = h_{ab} \delta g^{ab}.
\end{split}
\]
Therefore, we have 
\[
\delta \alpha = -\frac12 h_{ab} \delta g^{ab}
\]
and, similarly,
\[
\delta \beta = -\frac12 \tilde S_{ab}\delta g^{ab}.
\]
Now we can write down the variation of the energy 
\[
\delta \rho - \frac{\delta n}{n} \rho = n \left(A \alpha \delta \alpha
  + B \delta \beta \right) = -\frac{n}2 \left[ A \alpha h_{ab} + B
  \tilde S_{ab} \right] \delta g^{ab}.
\]
With this in hand we can now finally write down the energy-momentum
\[
T_{ab} = \rho\, u_a u_b  + n \alpha A h_{ab} + n B \tilde
  S_{ab}.
\]

\section{Energy-momentum tensors}
\label{sec:energy-moment-tens}

Using the abbreviations $x=F'e^{-(\xi-\xi_0)}$ and $y=F/r$ the
energy-momentum tensor for the Beig-Schmidt theory is obtained
from~(\ref{eq:BS_emt}) using the expression~(\ref{eq:eps}) for
$\eps_a{}^b$ in spherical symmetry. It has the following non-trivial
components 
\begin{gather}
  T_0{}^0 = \rho = xy^2 \left(\rho_0 + \frac\mu4\left\{ 
      (x^2-1)^2 + 2(y^2 - 1)^2\right\} + \frac\lambda8 \left\{(x^2-1) + 2 (y^2 -
      1) \right\}^2 \right),\label{eq:BST00}\\
  T_1{}^1 = -P = -\frac12 (xy^2)x^2 \left( (\lambda + 2\mu) (x^2 - 1) 
    + 2 \lambda (y^2 - 1) \right),\label{eq:BST11}\\
  T_2{}^2 = T_3{}^3 = -Q = -\frac12 (xy^2)y^2 \left((2\lambda + 2\mu)
    (y^2-1)  +  \lambda (x^2-1)\right).\label{eq:BST22}
\end{gather}
The energy-momentum for the Kijowski-Magli formulation in spherical
symmetry is obtained in a similar way from~(\ref{eq:KM_emt}) once the
tensor $S_a{}^b$ and its invariants $\alpha$ and $\beta$ are
determined. Since $S=-\frac12\log(1+\eps)$ and $\eps$ is diagonal in
spherical symmetry this is straightforward:
\begin{equation}
  \label{eq:S_in_spherical_symmetry}
  S_a{}^b \doteq -(\log x) dr
  \otimes \del_r - (\log y)\,
  (d\theta\otimes\del_\theta + d\phi\otimes \del_\phi).
\end{equation}
Hence, we get
\begin{equation}
  \label{eq:alpha}
  \alpha = -(\log x + 2 \log y),
\end{equation}
and 
\begin{equation}
  \label{eq:beta}
  \beta = \frac12\left[(\log x)^2 + 2 (\log y)^2\right] -\frac16 \left( \log
    x + 2 \log y \right)^2.
\end{equation}
With these expressions we obtain the non-trivial components of the
energy-momentum tensor
\begin{gather}
  T_0{}^0 = \rho = xy^2 \left(\rho_0 + \mu \left\{ 
      (\log x)^2 + 2(\log y)^2\right\} + \frac\lambda2
    \left\{\log x + 2 \log y\right\}^2 \right),\label{eq:KMT00}\\
  T_1{}^1 = -P = - (xy^2) \left( (\lambda + 2\mu)\log x 
    + 2 \lambda\log y \right),\label{eq:KMT11}\\
  T_2{}^2 = T_3{}^3 = -Q = - (xy^2) \left((2\lambda + 2\mu)
    \log y  +  \lambda \log x\right).\label{eq:KMT22}
\end{gather}

\section{Existence and Regularity of the solutions}
\label{sec:exist-regul-solut}

Our goal here is to see whether the theorem by Rendall and Schmidt can
be applied to our situation. For easy reference, we cite the theorem
here
\begin{theorem} [Rendall and Schmidt]
    Let $V$ be a finite dimensional vector space, $N: V\to V$ a
    linear map all of whose eigenvalues have positive real parts, and
    $ G: V \times (-\epsilon,\epsilon) \to V$ and $g:
    (-\epsilon,\epsilon) \to V$ smooth maps, where
    $\epsilon > 0$. Then, there exists $\delta< \epsilon$ and unique
    bounded $C^1$ function $u:(-\delta,0) \cup (0, \delta) \to V$
    which satisfies the equation
  \begin{equation}
    \label{eq:ode}
    s\frac{du}{ds}(s) + N u(s) = sG(s,u(s)) + g(s)
  \end{equation}
  Moreover, $u$ extends to a smooth solution of~(\ref{eq:ode}) on
  $(-\delta,\delta)$. If $N$, $G$ and $g$ depend smoothly on a
  parameter $t$ and the eigenvalues of $N$ are distinct, then the
  solution depends smoothly on $t$.
\end{theorem}
To analyze the behavior of the equations at the
center we follow the paper
by Park~\cite{park2000:static-solutions-of-the-einstein} who has found
that the equations for the Kijowski-Magli allow smooth regular
solutions. We need the same result for the equations coming from the
formulation of Beig and Schmidt.

It is easy to see that a regular spherically symmetric scalar function
$u(r)$ on a spherically symmetric space-time can smoothly
be extended as an even function of $r$ to negative values of the
radius. Thus, they depend smoothly on $s=r^2$ and their derivative at
the center vanishes. Using $s$ as the independent parameter instead of
$r$ our system~(\ref{eq:ode_system}) is
\begin{align}
  2sw' + 3 w&= 4\pi \rho,\\
  2sy' + y &= x f\\
  2sP_x x' &= P_y(y-f x) -s f^2(w+4\pi P) (P+\rho) - 2(P - Q),
\end{align}
where we have introduced $f=(1-2sw)^{-1/2}$. The solutions of this
system will be regular only if $x(0)=y(0)=a$, $P(0)=Q(0)$ and
$w(0)=4\pi/3 \rho(0)$. Thus, we can write
\[
x = a + s x_1(s), \qquad y = a + s y_1(s)
\]
and derive equations for the functions $x_1(s)$ and $y_1(s)$. This
yields the equations
\begin{align*}
  2sw'  + 3 w&= 4\pi \rho,\\
  2sy_1' + 3y_1 &= a \frac{f-1}{s} + x_1 f,\\
2s x_1' + 2 x_1 &=  \frac{P_y}{P_x}\left(\frac{1-f}{s} a + (y_1-f x_1)\right) -
\frac{f^2}{P_x}(w+4\pi P) (P+\rho) - \frac2{s}\frac{P-Q}{P_x}.
\end{align*}
The desired form is 
\[
2su' + Nu = sG(u,s) + g(s).
\]
The first equation is already in this form. In the second 
equation, the right hand side contains non-linear terms in $w$
(via $f$) but these terms have no factor $s$ in front. So we expand
\[
f = \left(1-2sw\right)^{-1/2} = 1 + sw + \cO(s^2)
\]
and get
\[
  2sy_1' + 3y_1 = a w + x_1 + \cO(s).
\]
Thus, this equation is in the desired form. Note, that it is not
important for this analysis to know the $\cO(s)$ exactly. It is enough
to know that they are well behaved at the origin. The $\cO(s)$ implies
that they have the desired factor of $s$ in front.

The third equation is more difficult.Here, the right hand side
of the equations contains non-linear terms in $w$, $x_1$ and $y_1$
through the dependence of $f$ and the functions $P$, $\rho$ and
$Q$. Therefore, we have to expand the right hand side around
$s=0$ and see whether we can isolate constant linear terms and
non-linear terms proportional to $s$. So we write
\[
P=P(x,y) = P(a+s x_1,a+s y_1) = P(a,a) + s (x_1P_x(a,a) + y_1
P_y(a,a)) + \cO(s^2)
\]
and similarly for all the other functions on the right hand side. We
write $P(a,a)=P^0$ and similarly for the other functions. First, we
observe that we need to have
\[
P^0-Q^0=0\qquad \text{and }\qquad P_x^0\ne0
\]
for the right hand side to be regular at all. Expanding the right hand
side as indicated we get
\[
2s x_1' + 4 x_1 
+ w \frac{P^0 + \rho^0 - a P_y^0}{P_x^0}
+ y_1 \frac{P_y^0 - 2 Q_y^0}{P_x^0}
+ x_1 \frac{P_y^0 - 2 Q_x^0}{P_x^0}
= -4\pi P^0 \frac{\rho^0 + P^0}{P_x^0} + \cO(s).
\]
This is the desired form and we can collect the coefficients of the linear
terms in the matrix $N$
\[
N = 
\begin{pmatrix}
3 & 0 & 0\\
-a & 3 & -1\\
  \frac{P^0 + \rho^0 - a P_y^0}{P_x^0}
   & \frac{P_y^0 - 2 Q_y^0}{P_x^0} & 4+\frac{P_y^0 - 2 Q_x^0}{P_x^0}
\end{pmatrix}.
\]
For the theorem to apply the eigenvalues of this matrix must be
positive. This leads to the following conditions: if $P_x^0>0$ then also
\begin{align}
  6 P_x^0 + 2 P_y^0 - 3 Q_x^0 - Q_y^0 &> 0,\\
  7 P_x^0 +   P_y^0 - 2 Q_x^0       &> 0,
\end{align}
while for $P_x^0<0$ the above expressions must also be negative. 

Now we have reduced the question of regularity at the center to a
condition for the energy functional which specifies the elastic
properties of the material. It is straightforward to check that these
conditions are satisfied for both the Beig-Schmidt and
Kijowski-Magli stored energy functionals.

\bibliography{Artikel}

\begin{thebibliography}{22}
\expandafter\ifx\csname natexlab\endcsname\relax\def\natexlab#1{#1}\fi
\expandafter\ifx\csname bibnamefont\endcsname\relax
  \def\bibnamefont#1{#1}\fi
\expandafter\ifx\csname bibfnamefont\endcsname\relax
  \def\bibfnamefont#1{#1}\fi
\expandafter\ifx\csname citenamefont\endcsname\relax
  \def\citenamefont#1{#1}\fi
\expandafter\ifx\csname url\endcsname\relax
  \def\url#1{\texttt{#1}}\fi
\expandafter\ifx\csname urlprefix\endcsname\relax\def\urlprefix{URL }\fi
\providecommand{\bibinfo}[2]{#2}
\providecommand{\eprint}[2][]{\url{#2}}

\bibitem[{\citenamefont{Hughes}(2000)}]{hughe2000:the-finite-element-method}
\bibinfo{author}{\bibfnamefont{T.~J.~R.} \bibnamefont{Hughes}},
  \emph{\bibinfo{title}{The finite element method}}
  (\bibinfo{publisher}{Dover}, \bibinfo{year}{2000}).

\bibitem[{\citenamefont{Synge}(1959)}]{synge1959:a-theory-of-elasticity-in-gen%
eral}
\bibinfo{author}{\bibfnamefont{J.~L.} \bibnamefont{Synge}},
  \bibinfo{journal}{Math. Z.} \textbf{\bibinfo{volume}{72}},
  \bibinfo{pages}{82} (\bibinfo{year}{1959}).

\bibitem[{\citenamefont{Rayner}(1963)}]{rayne1963:elasticity-in-general-relati%
vity}
\bibinfo{author}{\bibfnamefont{C.~B.} \bibnamefont{Rayner}},
  \bibinfo{journal}{Proc. Roy. Soc. London A} \textbf{\bibinfo{volume}{272}},
  \bibinfo{pages}{44} (\bibinfo{year}{1963}).

\bibitem[{\citenamefont{Carter and
  Quintana}(1972)}]{cartequint1972:foundations-of-general-relativistic}
\bibinfo{author}{\bibfnamefont{B.}~\bibnamefont{Carter}} \bibnamefont{and}
  \bibinfo{author}{\bibfnamefont{H.}~\bibnamefont{Quintana}},
  \bibinfo{journal}{Proc. Roy. Soc. London A} \textbf{\bibinfo{volume}{331}},
  \bibinfo{pages}{57} (\bibinfo{year}{1972}).

\bibitem[{\citenamefont{Kijowski and
  Magli}(1992)}]{kijowmagli1992:relat-elastomechanics}
\bibinfo{author}{\bibfnamefont{J.}~\bibnamefont{Kijowski}} \bibnamefont{and}
  \bibinfo{author}{\bibfnamefont{G.}~\bibnamefont{Magli}}, \bibinfo{journal}{J.
  Geom. Phys.} \textbf{\bibinfo{volume}{9}}, \bibinfo{pages}{207}
  (\bibinfo{year}{1992}).

\bibitem[{\citenamefont{Beig and Schmidt}(2003)}]{beigschmidt2003:_relat_elast}
\bibinfo{author}{\bibfnamefont{R.}~\bibnamefont{Beig}} \bibnamefont{and}
  \bibinfo{author}{\bibfnamefont{B.~G.} \bibnamefont{Schmidt}},
  \bibinfo{journal}{Class. Quant. Grav.} \textbf{\bibinfo{volume}{20}},
  \bibinfo{pages}{889} (\bibinfo{year}{2003}).

\bibitem[{\citenamefont{Beig and
  Schmidt}(2006)}]{beigschmi2006:relativistic-elastostatics-i}
\bibinfo{author}{\bibfnamefont{R.}~\bibnamefont{Beig}} \bibnamefont{and}
  \bibinfo{author}{\bibfnamefont{B.~G.} \bibnamefont{Schmidt}},
  \bibinfo{journal}{Classical and Quantum Gravity}  (\bibinfo{year}{2006}).

\bibitem[{\citenamefont{Andersson et~al.}(2006)\citenamefont{Andersson, Beig,
  and Schmidt}}]{anderbeig2006:static-self-gravitating-elastic}
\bibinfo{author}{\bibfnamefont{L.}~\bibnamefont{Andersson}},
  \bibinfo{author}{\bibfnamefont{R.}~\bibnamefont{Beig}}, \bibnamefont{and}
  \bibinfo{author}{\bibfnamefont{B.~G.} \bibnamefont{Schmidt}},
  \emph{\bibinfo{title}{Static self-gravitating elastic bodies in einstein
  gravity}} (\bibinfo{year}{2006}),
  \urlprefix\url{http://arxiv.org/abs/gr-qc/0611108}.

\bibitem[{\citenamefont{Losert}(2006)}]{loser2006:static-elastic-shells}
\bibinfo{author}{\bibfnamefont{C.~M.} \bibnamefont{Losert}},
  \emph{\bibinfo{title}{Static elastic shells in einsteinian and newtonian
  gravity}} (\bibinfo{year}{2006}),
  \urlprefix\url{http://arxiv.org/abs/gr-qc/0603103}.

\bibitem[{\citenamefont{Karlovini and
  Samuelsson}(2003)}]{karlosamue2003:elastic-stars-in-general}
\bibinfo{author}{\bibfnamefont{M.}~\bibnamefont{Karlovini}} \bibnamefont{and}
  \bibinfo{author}{\bibfnamefont{L.}~\bibnamefont{Samuelsson}},
  \bibinfo{journal}{Classical and Quantum Gravity}
  \textbf{\bibinfo{volume}{20}}, \bibinfo{pages}{3613} (\bibinfo{year}{2003}).

\bibitem[{\citenamefont{Karlovini et~al.}(2004)\citenamefont{Karlovini,
  Samuelsson, and Zarroug}}]{karlosamue2004:elastic-stars-in-general_II}
\bibinfo{author}{\bibfnamefont{M.}~\bibnamefont{Karlovini}},
  \bibinfo{author}{\bibfnamefont{L.}~\bibnamefont{Samuelsson}},
  \bibnamefont{and} \bibinfo{author}{\bibfnamefont{M.}~\bibnamefont{Zarroug}},
  \bibinfo{journal}{Classical and Quantum Gravity}
  \textbf{\bibinfo{volume}{21}}, \bibinfo{pages}{1559} (\bibinfo{year}{2004}).

\bibitem[{\citenamefont{Karlovini and
  Samuelsson}(2004)}]{karlosamue2004:elastic-stars-in-general_III}
\bibinfo{author}{\bibfnamefont{M.}~\bibnamefont{Karlovini}} \bibnamefont{and}
  \bibinfo{author}{\bibfnamefont{L.}~\bibnamefont{Samuelsson}},
  \bibinfo{journal}{Classical and Quantum Gravity}
  \textbf{\bibinfo{volume}{21}}, \bibinfo{pages}{4531} (\bibinfo{year}{2004}).

\bibitem[{\citenamefont{Szabados}(2004)}]{szabados2004:_quasi_local_energ_mome%
n_angul_momen_gr}
\bibinfo{author}{\bibfnamefont{L.~B.} \bibnamefont{Szabados}},
  \bibinfo{journal}{Living Rev. Relativity} \textbf{\bibinfo{volume}{7}}
  (\bibinfo{year}{2004}),
  \urlprefix\url{http://www.livingreviews.org/lrr-2004-4}.

\bibitem[{\citenamefont{Wald}(1984)}]{wald84:_gener_relat}
\bibinfo{author}{\bibfnamefont{R.~M.} \bibnamefont{Wald}},
  \emph{\bibinfo{title}{General Relativity}} (\bibinfo{publisher}{Chicago
  University Press}, \bibinfo{address}{Chicago}, \bibinfo{year}{1984}).

\bibitem[{\citenamefont{Rendall and
  Schmidt}(1991)}]{rendaschmi1991:existence-and-properties-of-spherically}
\bibinfo{author}{\bibfnamefont{A.}~\bibnamefont{Rendall}} \bibnamefont{and}
  \bibinfo{author}{\bibfnamefont{B.~G.} \bibnamefont{Schmidt}},
  \bibinfo{journal}{Class. Quant. Grav.} \textbf{\bibinfo{volume}{8}},
  \bibinfo{pages}{985} (\bibinfo{year}{1991}).

\bibitem[{\citenamefont{Haensel}(2001)}]{haens2001:neutron-star-crusts}
\bibinfo{author}{\bibfnamefont{P.}~\bibnamefont{Haensel}}, in
  \emph{\bibinfo{booktitle}{Physics of Neutron Star Interiors}}, edited by
  \bibinfo{editor}{\bibfnamefont{D.}~\bibnamefont{Blaschke}},
  \bibinfo{editor}{\bibfnamefont{N.}~\bibnamefont{Glendenning}},
  \bibnamefont{and} \bibinfo{editor}{\bibfnamefont{A.}~\bibnamefont{Sedrakian}}
  (\bibinfo{year}{2001}), vol. \bibinfo{volume}{578}, pp.
  \bibinfo{pages}{127--174}.

\bibitem[{\citenamefont{{Strohmayer} et~al.}(1991)\citenamefont{{Strohmayer},
  {van Horn}, {Ogata}, {Iyetomi}, and
  {Ichimaru}}}]{strohvan-h1991:shear-modulus-neutron}
\bibinfo{author}{\bibfnamefont{T.}~\bibnamefont{{Strohmayer}}},
  \bibinfo{author}{\bibfnamefont{H.~M.} \bibnamefont{{van Horn}}},
  \bibinfo{author}{\bibfnamefont{S.}~\bibnamefont{{Ogata}}},
  \bibinfo{author}{\bibfnamefont{H.}~\bibnamefont{{Iyetomi}}},
  \bibnamefont{and}
  \bibinfo{author}{\bibfnamefont{S.}~\bibnamefont{{Ichimaru}}},
  \bibinfo{journal}{Astrophysical Journal} \textbf{\bibinfo{volume}{375}},
  \bibinfo{pages}{379} (\bibinfo{year}{1991}).

\bibitem[{\citenamefont{Kabobel}(2001)}]{kabobel2001:_symmet_in_relat}
\bibinfo{author}{\bibfnamefont{A.}~\bibnamefont{Kabobel}}, Master's thesis,
  \bibinfo{school}{Universit\"at T\"ubingen} (\bibinfo{year}{2001}).

\bibitem[{\citenamefont{Luk{\'a}cs}(1977)}]{lukac1977:finite-homogeneous-relat%
ivistic}
\bibinfo{author}{\bibfnamefont{B.}~\bibnamefont{Luk{\'a}cs}},
  \bibinfo{journal}{Nuovo Cimento} \textbf{\bibinfo{volume}{B40}},
  \bibinfo{pages}{169} (\bibinfo{year}{1977}).

\bibitem[{\citenamefont{Carter}()}]{personalcommunication}
\bibinfo{author}{\bibfnamefont{B.}~\bibnamefont{Carter}},
  \bibinfo{note}{personal communication}.

\bibitem[{\citenamefont{Park}(2000)}]{park2000:static-solutions-of-the-einstei%
n}
\bibinfo{author}{\bibfnamefont{J.}~\bibnamefont{Park}}, \bibinfo{journal}{Gen.
  Rel. Grav.} \textbf{\bibinfo{volume}{32}}, \bibinfo{pages}{235}
  (\bibinfo{year}{2000}), \urlprefix\url{gr-qc/9810010}.

\bibitem[{\citenamefont{Penrose and
  Rindler}(1984)}]{penrose84:_spinor_spacet_i}
\bibinfo{author}{\bibfnamefont{R.}~\bibnamefont{Penrose}} \bibnamefont{and}
  \bibinfo{author}{\bibfnamefont{W.}~\bibnamefont{Rindler}},
  \emph{\bibinfo{title}{Spinors and Spacetime}}, vol.~\bibinfo{volume}{1}
  (\bibinfo{publisher}{Cambridge University Press},
  \bibinfo{address}{Cambridge}, \bibinfo{year}{1984}).

\end{thebibliography}

\end{document}